\documentclass[12pt,preprint]{aastex}
\begin{document}
\shortauthors{Pen,Van Waerbeke,Mellier}
\shorttitle{Lensing Correlation}
\title{Gravity and Non-gravity Modes in the VIRMOS-DESCART
Weak Lensing Survey \thanks{
Based on observations obtained at the
Canada-France-Hawaii Telescope (CFHT) which is operated by the National
Research Council of Canada (NRCC), the Institut des Sciences de l'Univers
(INSU) of the Centre National de la Recherche Scientifique (CNRS) and
the University of Hawaii (UH)}}
\author{Ue-Li Pen}
\affil{Canadian Institute for Theoretical Astrophysics,
60 St George St., Toronto, Ont. M5S 3H8, Canada;
pen@cita.utoronto.ca}
\author{Ludovic van Waerbeke}
\affil{
Institut d'Astrophysique de Paris, 98 bis, boulevard Arago, 75014
Paris, France; waerbeke@iap.fr}
\affil{Canadian Institute for Theoretical Astrophysics,
60 St George St., Toronto, Ont. M5S 3H8, Canada}
\author{Yannick Mellier}
\affil{
Institut d'Astrophysique de Paris, 98 bis, boulevard Arago, 75014
Paris, France;
mellier@iap.fr}
\affil{
Observatoire de Paris, DEMIRM, 61 avenue de l'Observatoire, 75014
Paris, France}
\newcommand{\etal}{{\it et al. }}
\newcommand{\beq}{\begin{equation}}
\newcommand{\eeq}{\end{equation}}
\newcommand{\myplus}{+}
\newcommand{\myminus}{-}

\begin{abstract}

We analyze the weak lensing data of the VIRMOS imaging survey using
projections (called $E$ and $B$-modes) of the two independents observed
correlation functions. The $E$-mode contains all the lensing signal,
while noise and systematics contribute equally to the $E$ and $B$ modes
provided that intrinsic alignment is negligible.  The mode separation
allows a measurement of the signal with a $\sqrt{2}$ smaller error bars,
and a separate channel to test for systematic errors.  We apply various
transformations, including a spherical harmonic space power spectrum
$C^E_l$ and $C^B_l$, which provides a direct measurement of the projected
dark matter distribution for $500 < l < 10^4$.

\end{abstract}

\section{Introduction}

The measurements of the alignment of distant galaxies from ground based data
(\citet{2000A&A...358...30V,2000Natur.405..143W,2000MNRAS.318..625B,
2000astro.ph..3338P,2001A&A...368..766M}) and from space (\citet{2001ApJ...552L..85R})
provided the first convincing evidence of gravitational lensing by large scale
structures.
In a more recent work, \citet{2001A&A...374..757V} measured the cosmic shear signal
using different statistics, and shown the remarkable agreement between
them, which demonstrated the gravitational lensing origin
of the signal. However an indication of a remaining systematic were found in
the aperture mass measurements. It could be either due to an imperfect Point
Spread Function correction, and/or to an intrinsic alignment of galaxies arising
from spin-spin correlation for instance (\citet{2000astro.ph.12336C}). Although
a robust and ultimate method to clean the measured lensing signal from contamination
would be the measurement of the cross-correlation between different source planes,
it is in principle possible to clean existing data using a curl-free/curl modes
decomposition (called $E$ and $B$ modes). The reason is that gravitational
lensing produces curl-free shear patterns only, while systematics and intrinsic
alignments contribute to the $B$ mode as well (\citet{2000astro.ph.12336C}).
This paper is
a tentative to measure these modes separately in the lensing signal
measured in the VIRMOS-Descart \footnote{http://terapix.iap.fr/Descart/}
weak lensing survey (\citet{2001A&A...374..757V})
and to establish the lensing origin of the signal on a quantitative base.

The first Section establishes some basic relations, and defines the mode
decomposition of the shear correlation function. The second Section shows the
mode measurement on integrated shear correlation functions which are trivially
related to windowed variances measured by previous authors. The last Section
shows the power spectrum measurement for the two modes. The theoretical background
of this work was essentially developed in \citet{2000astro.ph.12336C}
and will not be detailed here.

\section{Correlation Functions}
\label{sec:corr}

Weak lensing shear is a spin 2 polarization field, and can be
described by a traceless two by two matrix at every point,
\beq
{\bf \gamma}=\gamma_{ij}=\left(\begin{array}{cc}
	\gamma_1  & \gamma_2 \\
	\gamma_2  & -\gamma_1 \end{array} \right)
\eeq
This matrix has two degrees of freedom at every point, much
like a vector field.  In analogy to a vector field, there are two
coordinate invariant projections, a divergence and a curl.  The former
is called $E$-mode in analogy to the electric field, and the latter
called $B$-mode which is divergence free and analogous to a magnetic
field (\citet{1998PhRvD..58059dbZ,astrophsz})

The weak lensing shear can be expressed in terms of the mean
expectation of source ellipticities and alignments
\beq
\gamma_1=\langle\epsilon \cos(2\theta)\rangle, 
\ \ \ \gamma_2=\langle \epsilon\sin(2\theta)\rangle
\label{eqn:gamma}
\eeq
where $\theta$ is the angle between the major axis of the source
galaxy and the $x$ axis, and $\epsilon=(a-b)/(a+b)$ is determined
by the major axis length $a$ and minor axis length $b$. Following
\citet{1991ApJ...380....1M,1992ApJ...388..272K} we define the shear components
($\gamma_t,\gamma_r$) in 
the frame of the line connecting a pair of galaxies.
In equation
(\ref{eqn:gamma}) this corresponds to defining the galaxy alignment
$\theta$ in the frame of the pair separation.

To date, all shear correlation function analysis have been performed
from measurements of $\langle \gamma_t(x)\gamma_t(x+r)\rangle$
and $\langle\gamma_r(x)\gamma_r(x+r)\rangle$.
The spin-2 correlation function can then be decomposed into two coordinate
invariant components, $\xi_\myplus$ and $\xi_\myminus$:
\beq
\xi_\myplus(r)=\frac{1}{2}\langle \gamma_{ij}(x) \gamma_{ij}(x+r) \rangle
=\langle \gamma_t(x)\gamma_t(x+r)\rangle+\langle \gamma_r(x)\gamma_r(x+r)
\rangle.
\label{eqn:xip}
\eeq
\beq
\xi_\myminus(x)=\langle \gamma_t(x)\gamma_t(x+r)\rangle -\langle \gamma_r(x)
\gamma_r(x+r)\rangle.
\label{eqn:xim}
\eeq
To test for systematic errors in weak
lensing mass reconstruction, one usually rotates all galaxies by 45
degrees, upon which the weak lensing signal should disappear.
Under this rotation (\ref{eqn:xip}) is unchanged, while
(\ref{eqn:xim}) changes sign.

Weak lensing arises from a gravitational potential.  The statistics
of this scalar field can be described by a single correlation function,
which means there must exist
a degeneracy between the two correlators (\ref{eqn:xip},\ref{eqn:xim}).
\citet{2000astro.ph.12336C} have shown how to transform these two
correlators into a pure $E$-mode which contains all the lensing signal,
and a pure $B$-mode which should contain only noise. Any remaining systematic
effects is expected to contribute to both $E$ and $B$-mode: residuals in the Point
Spread Function corrections will contribute equally, while
the intrinsic alignment correlation of the galaxies is expected to
give a higher amplitude in $E$ than in $B$. How much higher is still debated
in the literature (\citet{2000astro.ph.12336C},\citet{2000ApJ...545..561C},
\citet{2001MNRAS.320L...7C},\citet{2000MNRAS.319..649H}).
However, a general agreement among the literature is that our survey is deep enough
so that intrinsic alignment has a negligible contribution (less than 10\%, see for
instance \citet{2000ApJ...543L.107P}).

Following \citet{2000astro.ph.12336C}, we define
\beq
\xi'(r)=\xi_\myminus(r)+4\int_r^\infty \frac{dr'}{r'} \xi_\myminus(r')
	-12r^2 \int_r^\infty \frac{dr'}{r'^3}\xi_\myminus(r').
\label{eqn:xipr}
\eeq
The $E$ and $B$-type correlators are given as
\beq
\xi^E(r)=\frac{\xi_\myplus(r)+\xi'(r)}{2}\ \ \ \ \ \ 
\xi^B(r)=\frac{\xi_\myplus(r)-\xi'(r)}{2}
\label{eqn:xieb}
\eeq
The sum is local, $\xi^E+\xi^B=\xi_\myplus$, while the difference depends on
the difference correlator $\xi_\myminus$ at larger radii.  Rotating all
galaxies by 45 degrees swaps $\xi^E$ and $\xi^B$, but leaves the sum
unchanged.  If we rotate only one galaxy in each pair by 45 degrees,
both E and B correlators should disappear.
In practice, we only know the correlators out to a finite
radius $r_1$.  The integral in (\ref{eqn:xipr}) which should go to
infinity must thus be truncated at $r_1$.  The integral from $r_1$ to
infinity results in two integration constants which affect the
correlation function at smaller radii.  The integration constant for
the last term in (\ref{eqn:xipr}) is weighted by $(r/r_1)^2$, and
should not affect short range correlations significantly.  The middle
term results in a straight constant $\bar{\xi}$, which can be added to
the $E$ correlator and subtracted from the $B$ correlator.  The
decomposition (\ref{eqn:xieb}) is degenerate under a change
$\xi^E(r)\rightarrow\xi^E(r)+\bar{\xi},\ \ \xi^B(r)
\rightarrow\xi^B(r)-\bar{\xi}$.  This degeneracy traces back to the
underlying map decomposition degeneracy.  A constant shear signal
generates a constant correlation function.  But this constant shear is
zero under differentiation, and cannot be classified unambiguously as
curl or div type.  In practice, we integrate as far as we have data,
and parameterize all lack of knowledge in terms of the integration
constant $\bar{\xi}$.  If one wishes to test the null hypothesis that
there are no $B$ modes as predicted by weak lensing, one can
marginalize the results of all possible integration constants. If
$B$ cannot be set to zero at all scales for any value of the
integration constant, we reject the hypothesis that there are no $B$
modes.  This would be an indicator of instrumental systematics.
Since there are two integration constants, one should marginalize over
both.  One of the constants only affects large scales, so at smaller
scales only $\bar{\xi}$ is of importance.

\newcommand{\ve}{\vec{e}}
\newcommand{\bN}{{\bf N}}

The raw correlation functions $\xi_\myplus,\ \xi_\myminus$ are measured by
averaging all pairs of galaxies at a given separation $r$.  In
practice, we binned the separations into intervals of three pixels,
corresponding to about 0.6 arc seconds, for a total of 10000 bins
corresponding to 1 2/3 degrees separation.  Each pair is given a
statistical weight defined by:

\beq
W_{ij}={1\over (\sigma_i^2+\sigma_e^2)}{1\over (\sigma_j^2+\sigma_e^2)},
\eeq
where $\sigma_i$ and $\sigma_j$ are the ellipticity measurement
r.m.s. of the galaxies $i$ and $j$. The quantity $\sigma_e$ is the
r.m.s ellipticity estimated over all the galaxies. This is
different from the weighting scheme expressed in Eq.(7) in
\citet{2000A&A...358...30V} but it gives similar results.
The sum of weights of pairs in each bin
provides a total weight $w$ for each bin.

A crucial step is now the calculation
of the noise covariance matrix. Since the number of pairs is the
square of the number of actual galaxies, one might expect the
statistical error of each correlation bin to be slightly correlated with other
bins. In practice, we have not been able to find such a correlation (because the
bins are small and the noise dominates over the signal),
we thus treat each correlation bin as uncorrelated with the others. 
Each of the correlators
$\xi_\myplus,\xi_\myminus$ now has mutually identical, independent and radially
uncorrelated bins.  The mapping (\ref{eqn:xipr}) correlates bins at
different separation, while the transformation (\ref{eqn:xieb}) now
correlates the $E$ and $B$ correlators.  We compute the noise covariance
matrix as follows. The variances are discretized as a one dimensional array
$\ve=\{\sigma^2(r_i)\}$.  We will use the subscript $i$ to index this
one dimensional bin array.  Since we used 10000 bins $r_i$ to compute the
correlation functions, we can define the 20000 elements vector
$\xi_i=(\xi_\myplus(r_i),\xi_\myminus(r_i))$.
The noise covariance matrix
$N_{ij}=\langle(\xi_i-\langle\xi_i\rangle)(\xi^t_j-\langle\xi^t_j\rangle)\rangle$ of the raw correlators is a $20000^2$ matrix which is $\ve$
on the diagonal and zero elsewhere.  The transformation into $E$ and
$B$ correlators from (\ref{eqn:xieb}) is a linear operation on the local correlators
$\xi_i$. We again define the 20000 elements vector $\xi^{EB}_i=(\xi^E(r_i),
\xi^B(r_i))$. Equation (\ref{eqn:xieb}) implicitly
relates them through a transformation
\beq
\xi^{EB}_i=T_{ij}\xi_j .
\label{eqn:ximat}
\eeq
Since each bin in $\xi^{EB}$ is extremely noisy, we have re-binned it
into 7 logarithmic intervals each a factor of two wide, denoted
$\xi^b_k$.  This re-binning can be represented as a 14 by 20000
projection operator $\xi^b_k=P_{ki}\xi^{EB}_i$.  The binned noise
covariance matrix $\bN^b$ is
\beq
N^b_{lm}=\langle (\xi^b_l-\langle\xi^b_l\rangle)(\xi^b_m-\langle\xi^b_m\rangle) \rangle = P_{li} T_{ij} N_{jk}
T^t_{ko} P^t_{om}.
\label{eqn:noiseb}
\eeq
\begin{figure}
\plotone{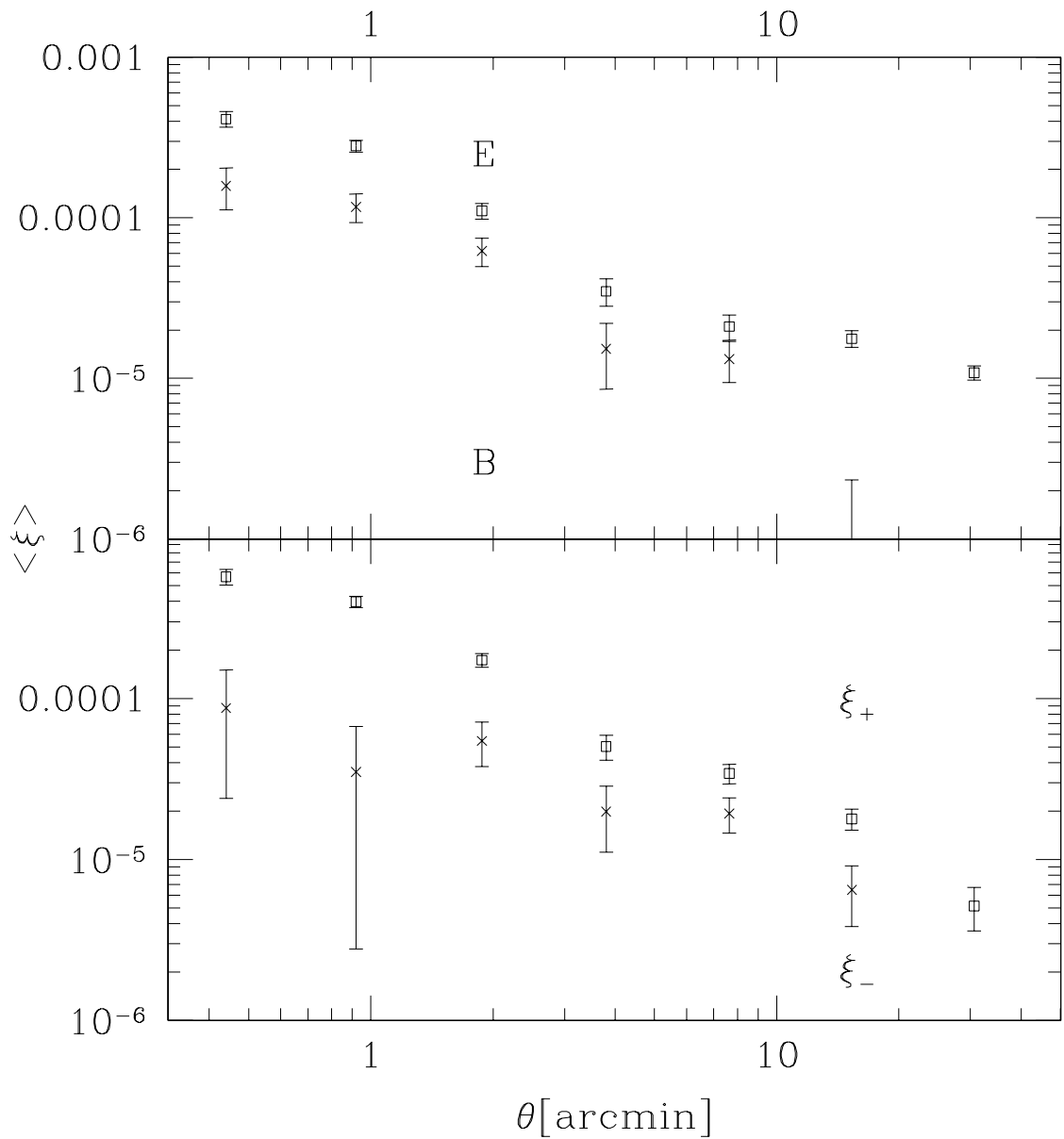}
\caption{Correlation functions.  The top panel shows the $E$-type
correlator (boxes) and the $B$-type correlation (crosses).  The bottom
panel shows the raw observables: the total correlation $\xi_\myplus$ (boxes)
and the differential correlation $\xi_\myminus$ (crosses).  The error
bars in the bottom panels are fully uncorrelated, while they are
mostly uncorrelated in the top panel.
}
\label{fig:corr}
\end{figure}
We show the result of the two projections in figure \ref{fig:corr}.
The error bars are the square root of the diagonal entries of $\bN^b$
(\ref{eqn:noiseb}), and correspond to one $\sigma$ error bars keeping
all other data points fixed.  We have normalized the degeneracy
$\bar{\xi}$ such that $\xi^B$ is zero at the bin at 15 arc minutes.
This second-to-largest scale was chosen for this normalization since
the largest bin has a significant dependence on the second integration
constant.

The $B$
correlator is systematically positive, indicating that some residual
non-lensing correlations may be contributing.  The $B$ mode is
about three times smaller than the $E$ mode, so this systematic is clearly
significantly weaker than the lensing signal.  Due to symmetry, the
$E$ and $B$ error bars are equal in magnitude, with the latter appearing
larger due to the logarithmic scale. However, the amplitude of the
$B$-mode is larger than the intrinsic alignment prediction which should not
exceed a few percents of the lensing signal here. Therefore we conclude that
most of the $B$-mode measured here is due to Point Spread Function and/or
image analysis residual error.

%
\begin{figure}
\plotone{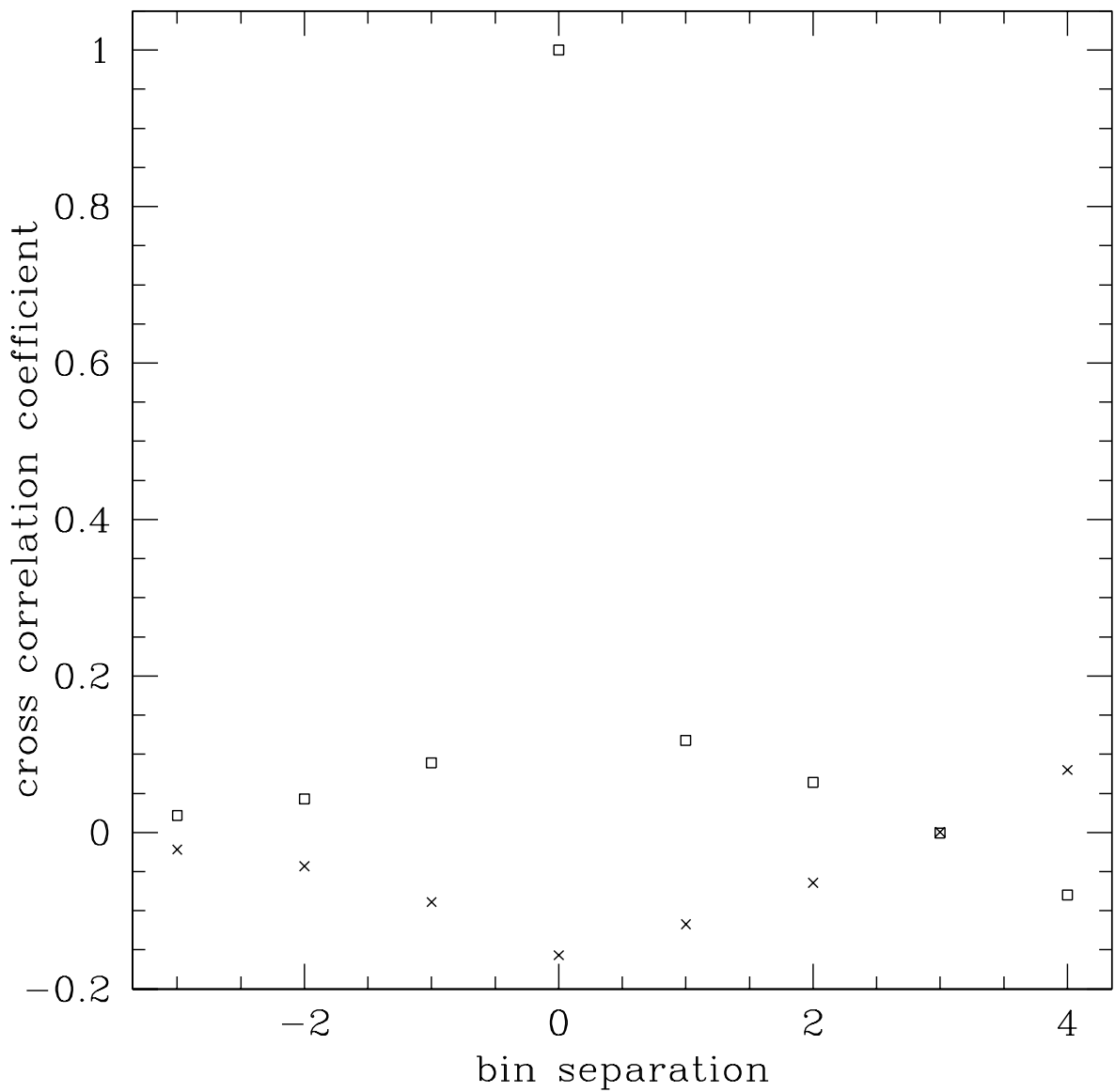}
\caption{Cross correlation coefficient.  The two panels show the
corresponding cross correlation coefficient between bins in figure
\protect\ref{fig:var}.  The boxes are the cross correlations for bins of
the same type (i.e. $E-E$ or $B-B$), and the crosses between types
(e.g. $E-B$).  All covariances measured relative to the bin at 4
arcminutes.  Other covariances are similar.  Not included is the
integration constant which can move one global offset from all $E$
bins to the $B$ bins.
}
\label{fig:coveb}
\end{figure}
The bin cross correlations are shown in figure \ref{fig:coveb}.  The
cross correlations are quite small, always less than 0.2, except for
the auto-correlation of each point, which by definition is 1.  We have
defined the cross correlation coefficient
\beq
c_{ij}=\frac{\langle \xi^b_i \xi^b_j \rangle }{\sqrt{\langle
(\xi^b_i)^2\rangle \langle (\xi^b_j)^2 \rangle}}.
\eeq
The $E$ and $B$ correlations are slightly anti-correlated.

\section{Windowed Variance}

From the two correlation functions $\xi_\myplus$ and $\xi_\myminus$, we can compute
the variances of the shear field smoothed with various windows.
\citet{2001A&A...374..757V} have measured the variances directly from
the smoothed maps.  Working on the maps is limited by geometry: they
have many defects which obliges us to mask a fair fraction ($\sim 20\%$) of
the observed area. These masks create many boundaries over the field,
which intersect quite often a smoothing window, enabling possible edge
effects. Integrating over the correlation function, this
problem does not exist.  \citet{2000astro.ph.12336C} have presented
various strategies to derive these variances from correlation
functions. 

\subsection{Top-Hat Filtering}
\begin{figure}
\plotone{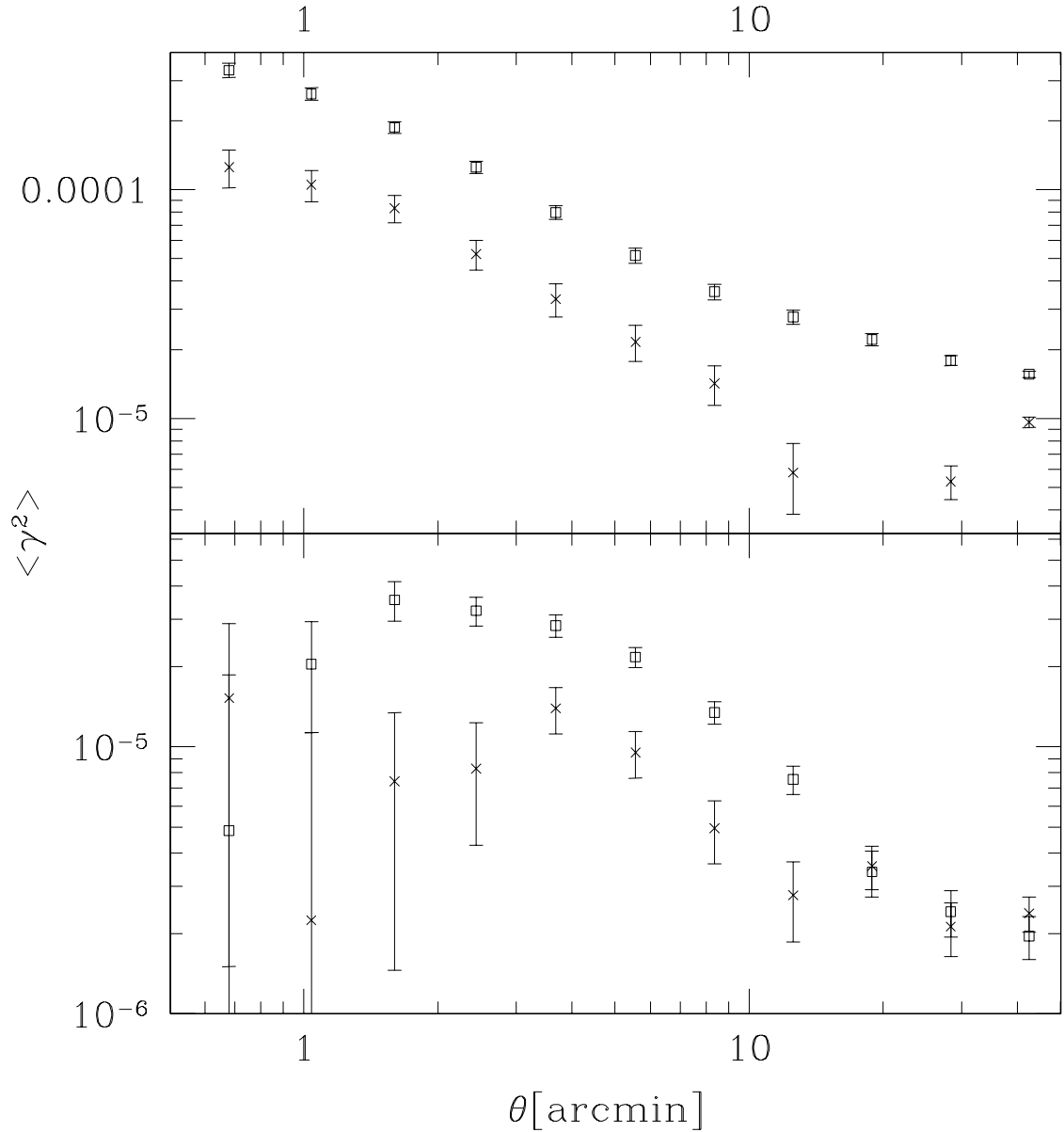}
\caption{Windowed Variances.  The top panel shows the top-hat variances
of the $E$-mode (boxes) and the $B$-mode (crosses).  The error bars
are highly correlated, with neighboring bins having a 90\%
cross-correlation coefficient.  The bottom panel shows the aperture
map variances, again with $E$-mode (boxes) and $B$-mode (crosses).
The bins are modestly correlated, with neighboring cross-correlation
coefficient of 50\%.}
\label{fig:var}
\end{figure}

\begin{figure}
\plotone{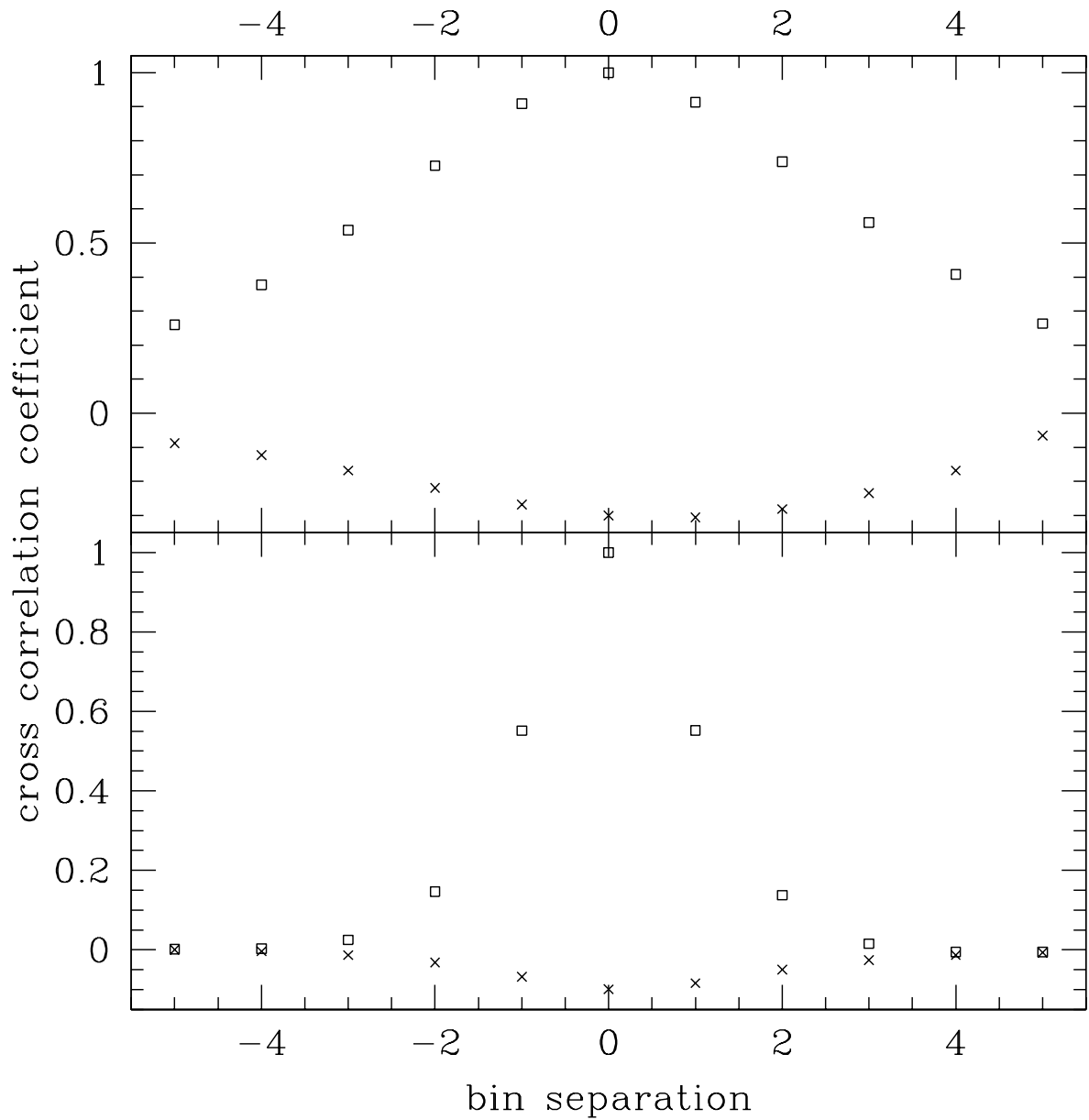}
\caption{Cross correlation coefficient.  The two panels show the
corresponding cross correlation coefficient between bins in figure
\protect\ref{fig:var}.  The boxes are the cross correlations for bins of
the same type (i.e. $E-E$ or $B-B$), and the crosses between types
(e.g. $E-B$).  All covariances measured relative to the bin at 6
arcminutes.  Other covariances are similar.  The top panel refers to
the top-hat, and the bottom panel to the aperture mass.
}
\label{fig:covar}
\end{figure}

The simplest statistic is the top-hat variance: one averages
the shear over a disk with some radius $R$, squares it and removes
the diagonal terms, such that we effectively measure a r.m.s. {\it excess} with
respect to random alignment of galaxies. We write the
smoothed shear field as
\beq
\gamma_i^R(x) = \int_{|x'|<R} \gamma_i(x-x') d^2x'.
\eeq
The expectation value of the variance of this smoothed field (which is
precisely the r.m.s. excess of galaxy ellipticities) can be
expressed in terms of the shear two point correlation function
$\xi_\myplus(r)$
\beq
\langle (\gamma_i^R)^2 \rangle = \frac{2}{\pi R^4}\int_0^{2R} rdr
\left[2R^2\cos^{-1}(\frac{r}{2R}) 
-r\sqrt{R^2-\frac{r^2}{4}}\right]\xi_\myplus(r)
\label{eqn:tophat}
\eeq
The top-hat variances published by previous authors was the sum of the $E$ and
$B$ type. Here,
again, we can decompose the correlation function $\xi_\myplus=\xi^E+\xi^B$, and
obtain two top-hat variance statistics. We show the results in
the top panel of Figure \ref{fig:var}.  The last two $B$ bins are
negative, and we took the absolute value to fit on this plot. 
The error bars are computed in the same way as in section
\ref{sec:corr}.
The covariance relative to the bin at 6 arcminutes is
shown in Figure \ref{fig:covar}. We see that the bins are highly
correlated, and there are only about
two ``independent'' degrees of freedom when summing over all radii.
The $B$ mode is a factor of 3
smaller than the $E$ mode, again suggesting that the lensing signal is
real.  This decomposition does depend on the integration constant
$\bar{\xi}$, which we have chosen to have zero $B$ mode at
19 arcmin. The
interpretation here would be that the data is almost consistent with
no $B$ mode, but does not rule out a large constant value which
cancels a large constant $E$ mode.  We discuss a statistic below
that does not have this degeneracy.

\subsection{Compensated Filtering}

The top-hat variance in (\ref{eqn:tophat}) is an integral measure of
the lensing power, and averages over all scales up to the smoothing
scale.  It has relatively small error bars since it bins together the
signal on many different scales.  It is thus not very sensitive to
measure scale dependence.  While the total top-hat power depends only
on galaxy pairs at separations up to twice the smoothing radius
(i.e. one smoothing diameter), the separation into $E$ and $B$ modes
is non-local, and depends on large scale correlations.

A purely local decomposition into $E$ and $B$ variances can be
achieved using aperture mass estimators \citep{1998MNRAS.296..873S}.
Locality here means that the decomposition into $E$- and $B$-type
variances on scale $R$ depends only on galaxy pairs whose separation
is less than $2R$.  We define a zero mean mass aperture window
\beq
{\cal U}(r)=\frac{9}{\pi}(1-r^2)\left(\frac{1}{3}-r^2\right)
\label{eqn:ur}
\eeq
for which $\int_0^1 rdr {\cal U}(r)=0$.  The aperture mass
\beq
\gamma_R=\frac{1}{R^2}\int d^2r \gamma_t({\bf r}) {\cal Q}(r/R)
=\frac{1}{R^2}\int d^2r \kappa({\bf r}) {\cal U}(r/R)
\label{eqn:grk}
\eeq
for the tangential shear $\gamma_t$ corresponding to $\gamma_1$ in the
frame of ${\bf r}$.  It expresses the aperture weighted optical depth
$\kappa$ in terms of the observable shear variation.  The shear window
is ${\cal Q}(r)={\cal U}(r)-\frac{2}{r^2}\int_0^r r'dr'{\cal U}(r')$.
These can be expressed as local integrals over the correlation
functions
\beq
\langle (\gamma^E_R)^2 \rangle
=\pi\int_0^{2R} rdr {\cal W}(r) \xi_\myplus(r) + \pi
\int_0^{2R}  rdr \tilde{\cal W}(r) \xi_\myminus(r).
\label{eqn:gape}
\eeq
${\cal W}$ is the 2-dimensional auto-convolution of ${\cal U}$ with
itself, and $\tilde{\cal W}$ is defined according to
\citet{2000astro.ph.12336C}.  The $B$ variance arises by using
$\gamma_2$ in (\ref{eqn:grk}), or changing the sign in the second term
of (\ref{eqn:gape}), which is zero for weak lensing.

We see the results in the lower panel Figure \ref{fig:var}.  There is
no integration constant, since the window has zero mean, so the $B$
mode is an independent estimate. For this reason, this plot can be
directly compared with Figure 4 in \citet{2001A&A...374..757V}: the results
are consistent but the errors are about twice smaller here, which means that
the small systematic which was measured in \citet{2001A&A...374..757V} is
in fact more
significant than what we expected. This mean that some among of signal cleaning
is required when estimating the cosmological parameters with such small error
bars (this will be done in a forthcoming paper), and also that it is necessary
to investigate in more details the possible source of systematics in the shape
measurement process.

\section{Power Spectrum}

We can measure the power spectrum directly from the correlation
function \citep{1998MNRAS.301.1064K}.  The Fourier space power spectra
trivially project the power into $E$ modes, which are aligned with the
wave vector, and $B$ modes which are at 45 degrees.
We have
\beq
P^E(k)=\pi \int_0^\infty \theta d\theta [\xi_\myplus J_0(k\theta)+\xi_\myminus J_4(k\theta)]
,\ \ \ \
P^B(k)=\pi \int_0^\infty \theta d\theta [\xi_\myplus J_0(k\theta)-\xi_\myminus J_4(k\theta)]
\label{eqn:pe}
\eeq
in terms of the Bessel functions $J_n$.  We map $k\sim l$ in
spherical harmonic space, and in the small angle approximation
$l(l+1) C_l/2\pi = k^2 P(k)/2\pi$.
The inverse mapping is given as
\beq
\xi_\myplus(\theta)=\frac{1}{2\pi}\int_0^\infty kdk [P^E(k)+P^B(k)] J_0(k\theta)
,\ \ \ \
\xi_\myminus(\theta)=\frac{1}{2\pi}\int_0^\infty kdk [P^E(k)-P^B(k)] J_4(k\theta)
\label{eqn:xipm}
\eeq

Numerically, the integral of (\ref{eqn:pe}) is dominated by angular
scales $k\sim 1/\theta$, but the noise may receive contributions from all
scales, especially scales for which there is little data and therefore
a lot of noise.  To minimize these effects, we used a maximum likelihood
power spectrum inversion.
In analogy to (\ref{eqn:ximat}) we can express (\ref{eqn:xipm}) as a linear
system
\beq
\xi^{\myplus\myminus}_i=T_{ij} P^{EB}_j \pm \sigma_i
\label{eqn:xipmeb}
\eeq
We re-scale (\ref{eqn:xipmeb}) by a diagonal noise matrix $N_{ii}=\sigma_i$
and parametrize $P^{EB}$ as a sum of band powers.  We used a sum
of 8 basis functions which is proportional to $1/k$ within each band.
$T_{ij}$ is then a 20000 by 16 matrix, and we obtain an over-determined
set of 20000 equations of equal weight for 16 unknowns.  The least
squares solution gives
\beq
P^{EB} =  \left( T^t N^t N T \right)^{-1} N^{-1} \xi^{\myplus\myminus}
\label{eqn:ls}
\eeq
The correlation function at zero lag is not observable, so the
determined angular power spectra are indeterminate up to constant.
The intrinsic ellipticity white noise has equal E and B modes, so
we are free to subtract the same constant white noise component from both.
We did so by forcing the B mode to be zero at $l\sim 5000$.  This
freedom of integration is not related to the integration constant in
the E-B separation (\ref{eqn:xipr}), which is one at large scales.  In
the latter case, we chose the integration from infinity rather than
integrating from zero.

The non-observability of the correlation function at zero lag is
fundamental.  $\xi(0)=\langle e^2\rangle+\langle\gamma^2\rangle$, but
we cannot measure the two terms separately.  In practice, the first
term, corresponding to intrinsic ellipticies, is much larger than the
rms gravitational shear described by the second term.  It is perhaps
amusing that the weak lensing power spectrum has an unobservable
offset, while the raw correlation functions $\xi^{\myplus\myminus}$ do not, in
contrast to galaxy power spectra where the converse is true.  One does
obtain lower bounds on the offset since power spectra cannot be
negative.  We chose the bin at $l\sim 5000$ since the B-mode had the
smallest relative error bar.  This subtraction leaves the higher $l$
slightly negative, but consistent with zero within the error bars of
each bin.  The indeterminacy of the zero lag correlation function also
manifests itself as an arbitrary integration constant $c/r^2$ falling
off as inverse separation squared in the aperture mass statistic,
i.e. a line of slope -2 in figure \ref{fig:var}.

\begin{figure}
\plotone{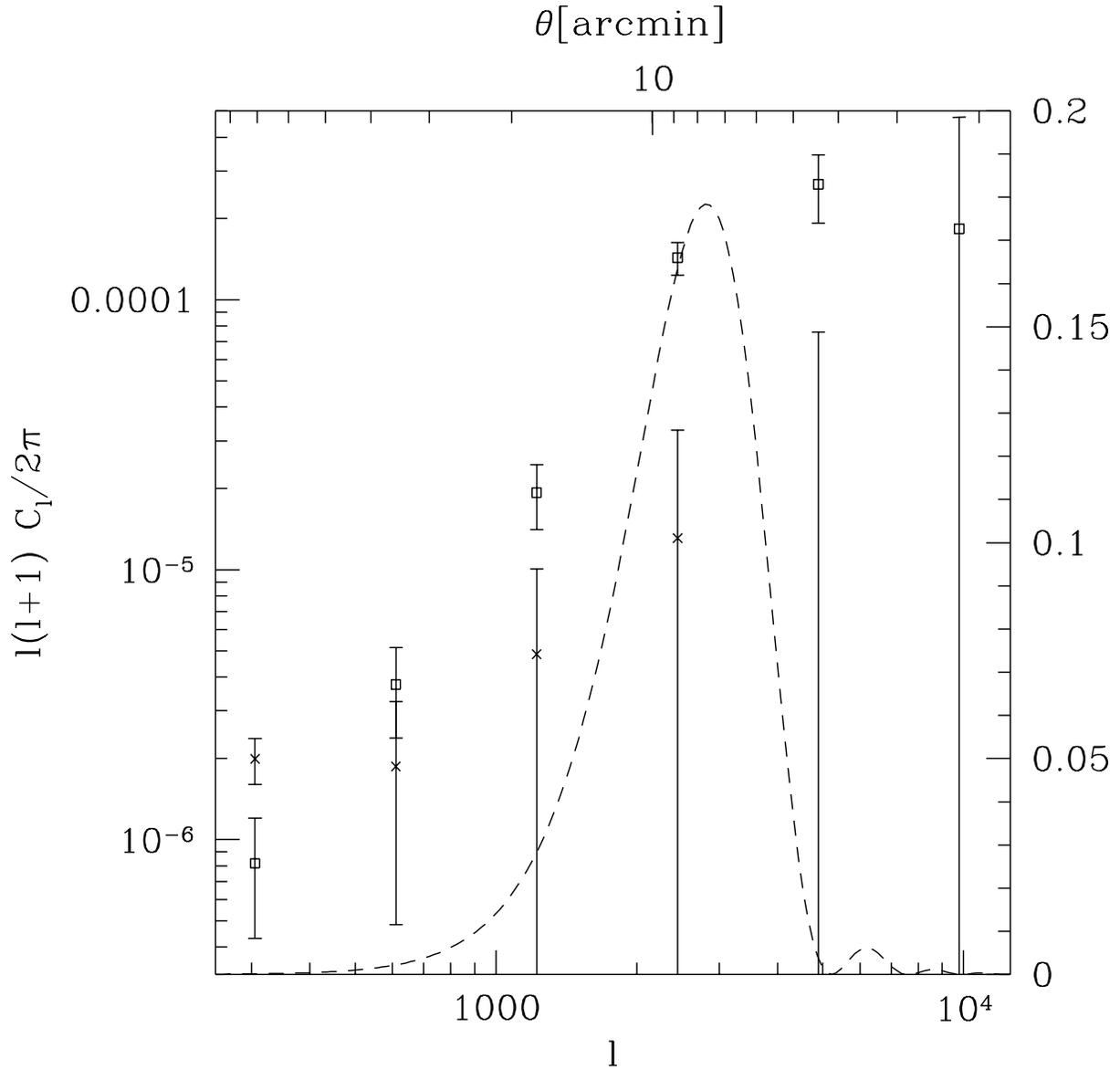}
\caption{Spherical harmonic power spectrum.  Again, the boxes indicate
the $E$ type power spectrum, while the crosses denote the $B$ type
power. The angular scale is the spacing of peaks of the wave
functions at the equator for m=l.  The dashed line is the window
function for the aperture mass statistic at 5', with amplitude
labeled on the right bounding box edge.  The aperture mass is
analogous to the integral of the window with the power spectrum using
equal weight per natural logarithmic interval.
}
\label{fig:cl}
\end{figure}

The result is shown in Figure \ref{fig:cl}. For the bin at 4.5arcmin, the
$B$-mode is consistent with zero. At large and small scales, we
have a larger amount of systematics.
This error analysis does not contain cosmic
variance, so this plot is not appropriate for measuring the slope of
the power spectrum.
\begin{figure}
\plotone{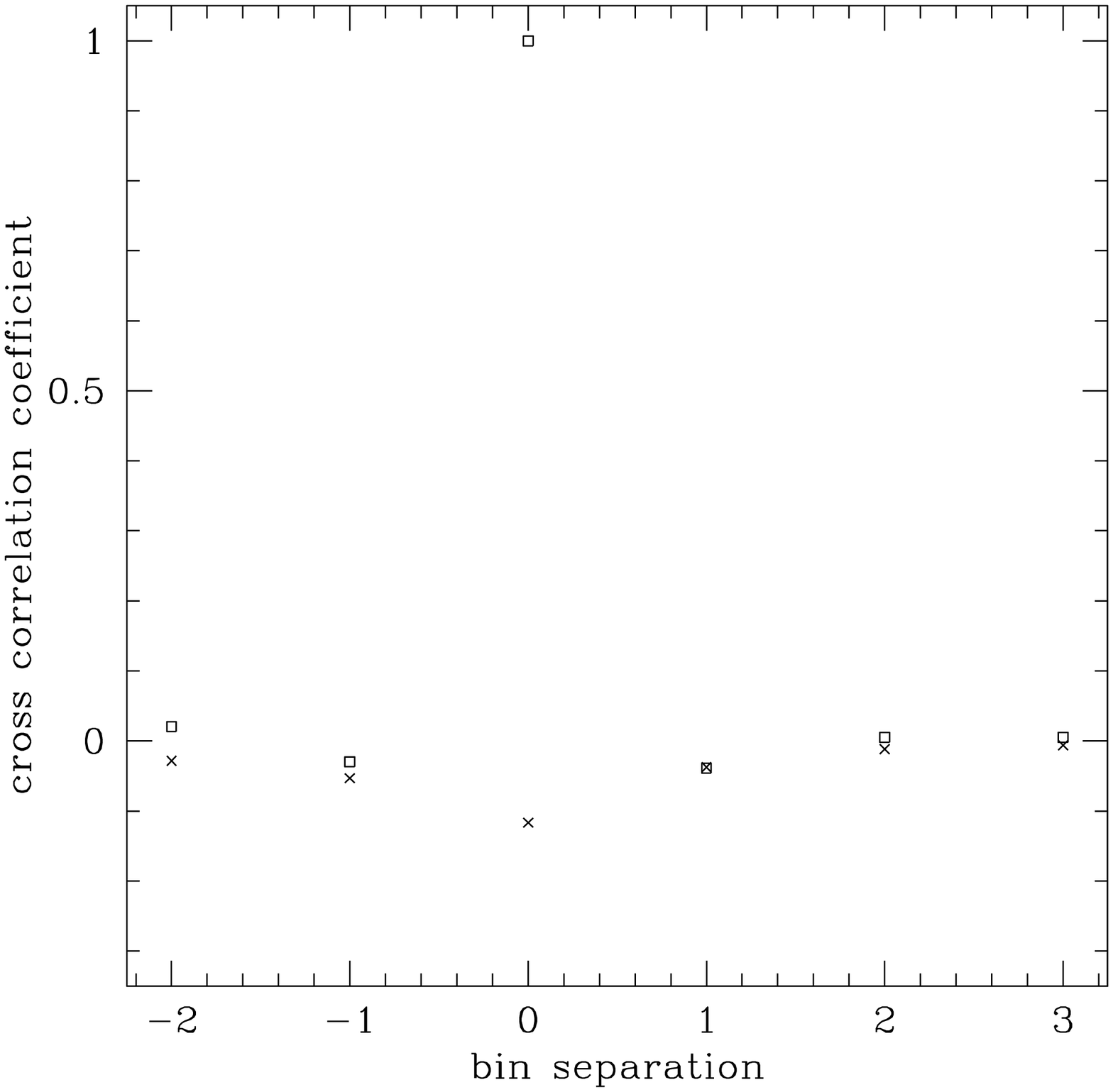}
\caption{Bin-bin cross correlation of the angular power spectrum in
Figure \protect\ref{fig:cl}.  The covariances are relative to the 
bin at l=1200.
}
\label{fig:cvcl}
\end{figure}
The covariance between bins is shown in Figure \ref{fig:cvcl},
showing a small covariance.

We introduce a mixing between modes since the integral (\ref{eqn:pe})
is truncated at both large and small radii.  This coupling is not
included in the error covariance.  To quantify this effect, we
generated a pure power law correlation function with the same binning.
We used $\xi_\myplus(\theta)=\xi_\myminus(\theta)\propto 1/\theta$, which is a pure
$E$ mode \citep{2000astro.ph.12336C}.  This test generated a small
negative $B$-mode which was largest at the leftmost bin, which was a
factor of 17 smaller in amplitude than the $E$-mode, and thus
insignificant compared to the other sources of error.

To compare the power spectrum analysis with the aperture mass window
from the previous section, we have overplotted the square of the
fourier transform of the aperture window (\ref{eqn:ur}) for a radius
of 5 arcmin in figure \ref{fig:cl} as a dashed line, with the window
units on the right box side.  This window peaks at a scale
corresponding to waves slightly longer than 5 arc minutes.  The window
has a minimum at $\theta_m=\pm \sqrt{2/3}\theta \sim 4'$ which has a
spacing of 8' between the two minima, comparable to a wave of
wavelength 8'.  We can see the qualitative drop of the aperture mass
near 20' in Figure \ref{fig:var} as the corresponding drop in the
power spectrum, where the E mode drops to the level of the B mode.

The optimal least squares inversion here has ignored the cosmic
variance, and is optimal in the limit that the signal to noise is very
small.  For the B mode, this is true at all scales, but for the E mode,
this procedure is not entirely optimal.  A proper power spectrum
estimation can be done by applying a full maximum likelihood estimate
to the data, which will be presented in a future paper.

\section{Conclusions}

We have derived several statistics from the two correlation functions
measured in the VIRMOS-Descart survey.  For each statistic, we are able to
decompose it into orthogonal $E$ and $B$ channels.  Since the two
channels add to the total power used in previous studies, the error in
each is $\sim \sqrt{2}$ smaller than in the total power.  The weak
lensing signal is solely contained in the $E$ channel, and the $B$
channel provides a monitor to watch for any potential systematic
contamination.  We have analyzed three categories of statistics: the
correlation function, windowed variances, and angular power spectrum.
In each case the $B$-mode was small, but statistically
significant, given the dramatic
reduction in the error bars compared to the previous analysis. The power
spectrum analysis
suggested that most of the systematics are coming from a constant value
$C_l$ which corresponds to a Poisson noise contribution. The origin of the
systematics is still unclear, but here we developed the tools useful for a
more detailed analysis, when the amount of data will be large enough.
Given that the $E$, $B$ correlation functions and the top-hat smoothed variances
are crucially dependent on the choice of an arbitrary integration constant,
the most robust statistic appears to be the aperture mass.
Our analysis shows for the first time a direct measurement of the projected
dark matter power spectrum, which opens a new window, not only for measuring
the cosmological parameters, but also for a future direct measure of the
three-dimensional mass power spectrum at small and intermediate scales.

{ \acknowledgements  We thank Dmitri Pogosyan and Simon Prunet for
useful discussions, and the VIRMOS and Terapix teams who got and processed the
VIRMOS-DESCART data. This work has been supported in part by NSERC
grant 72013704 and the CFI Pscinet computational resources, and by
the TMR Network ``Gravitational Lensing: New Constraints on
Cosmology and the Distribution of Dark Matter'' of the European Community
under contract No. ERBFMRX-CT97-0172.}

\appendix


\begin{thebibliography}{17}
\expandafter\ifx\csname natexlab\endcsname\relax\def\natexlab#1{#1}\fi

\bibitem[{{Bacon} {et~al.}(2000){Bacon}, {Refregier}, \&
  {Ellis}}]{2000MNRAS.318..625B}
{Bacon}, D.~J., {Refregier}, A.~R., \& {Ellis}, R.~S. 2000, \mnras, 318, 625

\bibitem[{{Catelan} {et~al.}(2001){Catelan}, {Kamionkowski}, \&
  {Blandford}}]{2001MNRAS.320L...7C}
{Catelan}, P., {Kamionkowski}, M., \& {Blandford}, R.~D. 2001, \mnras, 320, L7

\bibitem[{{Crittenden} {et~al.}(2000){Crittenden}, {Natarajan}, {Pen}, \&
  {Theuns}}]{2000astro.ph.12336C}
{Crittenden}, R.~G., {Natarajan}, P., {Pen}, U., \& {Theuns}, T. 2000, in 9
  pages, 3 postscript figures. Preprint no. DAMTP-2000-135., 12336+

\bibitem[{{Croft} \& {Metzler}(2000)}]{2000ApJ...545..561C}
{Croft}, R.~A.~C. \& {Metzler}, C.~A. 2000, \apj, 545, 561

\bibitem[{{Heavens} {et~al.}(2000){Heavens}, {Refregier}, \&
  {Heymans}}]{2000MNRAS.319..649H}
{Heavens}, A., {Refregier}, A., \& {Heymans}, C. 2000, \mnras, 319, 649

\bibitem[{{Kaiser}(1992)}]{1992ApJ...388..272K}
{Kaiser}, N. 1992, \apj, 388, 272

\bibitem[{{Kaiser} {et~al.}(2000){Kaiser}, {Wilson}, \&
  {Luppino}}]{2000astro.ph..3338P}
{Kaiser}, N., {Wilson}, G., \& {Luppino}, G. 2000, astro-ph/0003338 submitted
  to the ApJ Letters.

\bibitem[{{Kamionkowski} {et~al.}(1998){Kamionkowski}, {Babul}, {Cress}, \&
  {Refregier}}]{1998MNRAS.301.1064K}
{Kamionkowski}, M., {Babul}, A., {Cress}, C.~M., \& {Refregier}, A. 1998,
  \mnras, 301, 1064

\bibitem[{{Maoli} {et~al.}(2001){Maoli}, {Van Waerbeke}, {Mellier},
  {Schneider}, {Jain}, {Bernardeau}, {Erben}, \& {Fort}}]{2001A&A...368..766M}
{Maoli}, R., {Van Waerbeke}, L., {Mellier}, Y., {Schneider}, P., {Jain}, B.,
  {Bernardeau}, F., {Erben}, T., \& {Fort}, B. 2001, \aap, 368, 766

\bibitem[{{Miralda-Escud\'e}(1991)}]{1991ApJ...380....1M}
{Miralda-Escud\'e}, J. 1991, \apj, 380, 1

\bibitem[{{Pen} {et~al.}(2000){Pen}, {Lee}, \& {Seljak}}]{2000ApJ...543L.107P}
{Pen}, U., {Lee}, J., \& {Seljak}, U.~. 2000, \apjl, 543, L107

\bibitem[{{Rhodes} {et~al.}(2001){Rhodes}, {Refregier}, \&
  {Groth}}]{2001ApJ...552L..85R}
{Rhodes}, J., {Refregier}, A., \& {Groth}, E.~J. 2001, \apjl, 552, L85

\bibitem[{{Schneider} {et~al.}(1998){Schneider}, {van Waerbeke}, {Jain}, \&
  {Kruse}}]{1998MNRAS.296..873S}
{Schneider}, P., {van Waerbeke}, L., {Jain}, B., \& {Kruse}, G. 1998, \mnras,
  296, 873

\bibitem[{{Seljak} \& {Zaldarriaga}(1998)}]{astrophsz}
{Seljak}, U. \& {Zaldarriaga}, M. 1998, astro-ph/9805010, proceedings of the 
XXXIIIrd Rencontres de Moriond 1998.

\bibitem[{{Van Waerbeke} {et~al.}(2000){Van Waerbeke}, {Mellier}, {Erben},
  {Cuillandre}, {Bernardeau}, {Maoli}, {Bertin}, {Mc Cracken}, {Le F{\`e}vre},
  {Fort}, {Dantel-Fort}, {Jain}, \& {Schneider}}]{2000A&A...358...30V}
{Van Waerbeke}, L., {Mellier}, Y., {Erben}, T., {Cuillandre}, J.~C.,
  {Bernardeau}, F., {Maoli}, R., {Bertin}, E., {Mc Cracken}, H.~J., {Le
  F{\`e}vre}, O., {Fort}, B., {Dantel-Fort}, M., {Jain}, B., \& {Schneider}, P.
  2000, \aap, 358, 30

\bibitem[{{Van Waerbeke} {et~al.}(2001){Van Waerbeke}, {Mellier}, {Radovich},
  {Bertin}, {Dantel-Fort}, {McCracken}, {Le F{\` e}vre}, {Foucaud},
  {Cuillandre}, {Erben}, {Jain}, {Schneider}, {Bernardeau}, \&
  {Fort}}]{2001A&A...374..757V}
{Van Waerbeke}, L., {Mellier}, Y., {Radovich}, M., {Bertin}, E., {Dantel-Fort},
  M., {McCracken}, H.~J., {Le F{\` e}vre}, O., {Foucaud}, S., {Cuillandre},
  J.-C., {Erben}, T., {Jain}, B., {Schneider}, P., {Bernardeau}, F., \& {Fort},
  B. 2001, \aap, 374, 757

\bibitem[{{Wittman} {et~al.}(2000){Wittman}, {Tyson}, {Kirkman},
  {Dell'Antonio}, \& {Bernstein}}]{2000Natur.405..143W}
{Wittman}, D.~M., {Tyson}, J.~A., {Kirkman}, D., {Dell'Antonio}, I., \&
  {Bernstein}, G. 2000, \nat, 405, 143

\bibitem[{{Zaldarriaga} \& {Seljak}(1998)}]{1998PhRvD..58059dbZ}
{Zaldarriaga}, M. \& {Seljak}, U. 1998, \prd, 58, 023003 

\end{thebibliography}
\end{document}